\documentclass{iopconfser}

\usepackage{graphicx}
\usepackage{indentfirst}
\usepackage{amsmath}

\begin{document}

\title{Applicability evaluation of selected xAI methods for machine
learning algorithms for signal parameters extraction}

\author{Kalina Dimitrova$^{1}$, Venelin Kozhuharov$^{1}$ and Peicho Petkov$^{1}$}

\affil{$^1$Faculty of Physics, Sofia University ``St. Kliment Ohridski'', 5 J. Bourchier Blvd., 1164 Sofia, Bulgaria}

\email{kalina@phys.uni-sofia.bg}

\begin{abstract}
Machine learning methods find growing application in the reconstruction and analysis
of data in high energy physics experiments. A modified convolutional autoencoder model
was employed to identify and reconstruct the pulses from scintillating crystals. The model
was further investigated using four xAI methods for deeper understanding of the underlying reconstruction mechanism. The results are discussed in detail,
underlining the importance of xAI for knowledge gain and further improvement of the
algorithms.
\end{abstract}

\section{Introduction}
Machine learning (ML) has become increasingly important in high-energy physics (HEP) and nuclear physics research. It has been employed in diverse areas of data analysis, such as event selection \cite{Whiteson2009}, feature extraction in heavy-ion collisions \cite{Pang2021}, and the search for new fundamental physics \cite{Karagiorgi2022}. Deep learning techniques, in particular, have demonstrated significant potential in managing the intricate, high-dimensional data produced by the Large Hadron Collider \cite{Guest2018}. ML applications extend to the study of QCD matter under extreme conditions \cite{Zhou2024}. The integration of ML into high-energy and nuclear physics has facilitated advancements in understanding the nuclear structure, properties of the quark-gluon plasma, and the nuclear equation of state \cite{Pang2024}. As the field advances, there is a growing emphasis on incorporating physics-informed priors into ML models and addressing the challenges of applying these methodologies to complex physical systems \cite{He2023}.


Explainable AI (xAI) has emerged as a critical field addressing the need for transparency in complex AI systems \cite{Yadav2024, Zodage2024}. xAI aims to make AI processes more understandable and transparent, thereby enhancing trust, accountability, and collaboration between humans and AI \cite{Kamath2021, Kumawat2024}. Its importance is especially evident in high-stakes domains such as healthcare, finance, and legal systems \cite{Sewada2023}. xAI techniques encompass interpretable models and post-hoc explanations, such as LIME and SHAP \cite{Zodage2024, Chia2023}. These methods are instrumental in detecting biases, improving model performance, and ensuring ethical AI design \cite{Yadav2024, Tiwari2023}. Nevertheless, challenges persist in balancing interpretability with accuracy and in developing more efficient attribution methods \cite{Sewada2023, Adhikari2023}. 
xAI techniques for deep neural networks were initially applied to image processing models. Saliency maps are a widely used approach, with methods such as SmoothGrad \cite{Smilkov2017} and Occlusion Sensitivity Analysis \cite{Valois2023} enhancing basic gradient-based techniques. These methods aim to identify influential pixels or regions in input images \cite{Mundhenk2019}. SmoothGrad, for instance, adds noise to sharpen sensitivity maps, but it does not smooth the gradient \cite{Seo2018}. Various methods, including perturbation-based and gradient-based approaches, have been compared for their effectiveness in interpreting GeoAI models \cite{Hsu2023}. To evaluate explanation quality, measures such as infidelity and sensitivity have been proposed \cite{Yeh2019}. Despite these advancements, determining the optimal explanation method remains challenging, as different approaches like Gradient, Grad-CAM, and DEEPCOVER offer distinct advantages \cite{Konate2021}. Overall, these techniques have significantly improved the interpretability of deep learning models across diverse applications.


Explainable AI is increasingly important in high-energy physics (HEP) for interpreting complex neural networks and machine learning models. Various xAI methods, such as SHAP, are being applied to understand model decisions in particle classification and event reconstruction \cite{Pezoa2023, Roy2022}. These techniques help uncover feature importance, optimize models, and provide human-readable explanations \cite{Maglianella2023, Neubauer2022}. xAI is crucial for ensuring reliability and completeness in physics discoveries, particularly in tasks such as identifying Higgs boson decays and reconstructing particle positions in liquid argon detectors \cite{Turvill2020, Cardenas2024}. Additionally, xAI approaches are being explored to infer underlying physics from collision data and improve parton shower simulations \cite{Lai2020}. As AI and ML tools become more prevalent in theoretical particle physics, physics-aware development remains essential to ensure unbiased results and maintain a deep understanding of the underlying physics \cite{Gupta2022}.

In this study, we 
tested
xAI techniques such as 
Integrated gradients, 
Vanilla saliency, SmoothGrad, and Occlusion sensitivity to ML models for signal parameters extraction. 
Based on the xAI output, we have improved the understanding of the 
performance of the developed ML models and we assessed the possibility 
to further improve the models themselves.

\section{Machine learning models for signal parameters extraction}

ML techniques can be used as a method for event reconstruction in high-energy physics experiments. Such methods find growing application in the reconstruction of signals in electromagnetic calorimeters where the high particle multiplicities may lead to event pile-ups. This requires reconstruction methods with good double-pulse separation abilities and time resolution.

\begin{figure}[h!]
\centering
\includegraphics[width=\columnwidth]{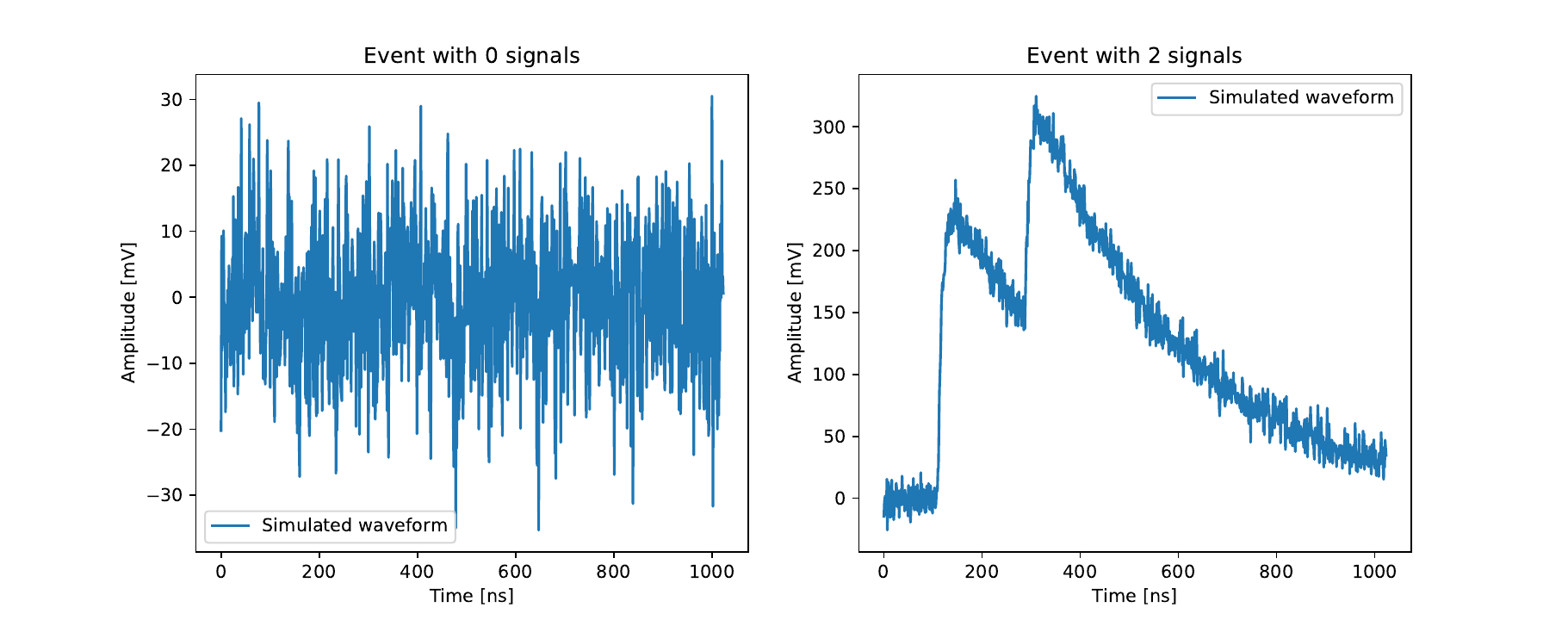}
\includegraphics[width=\columnwidth]{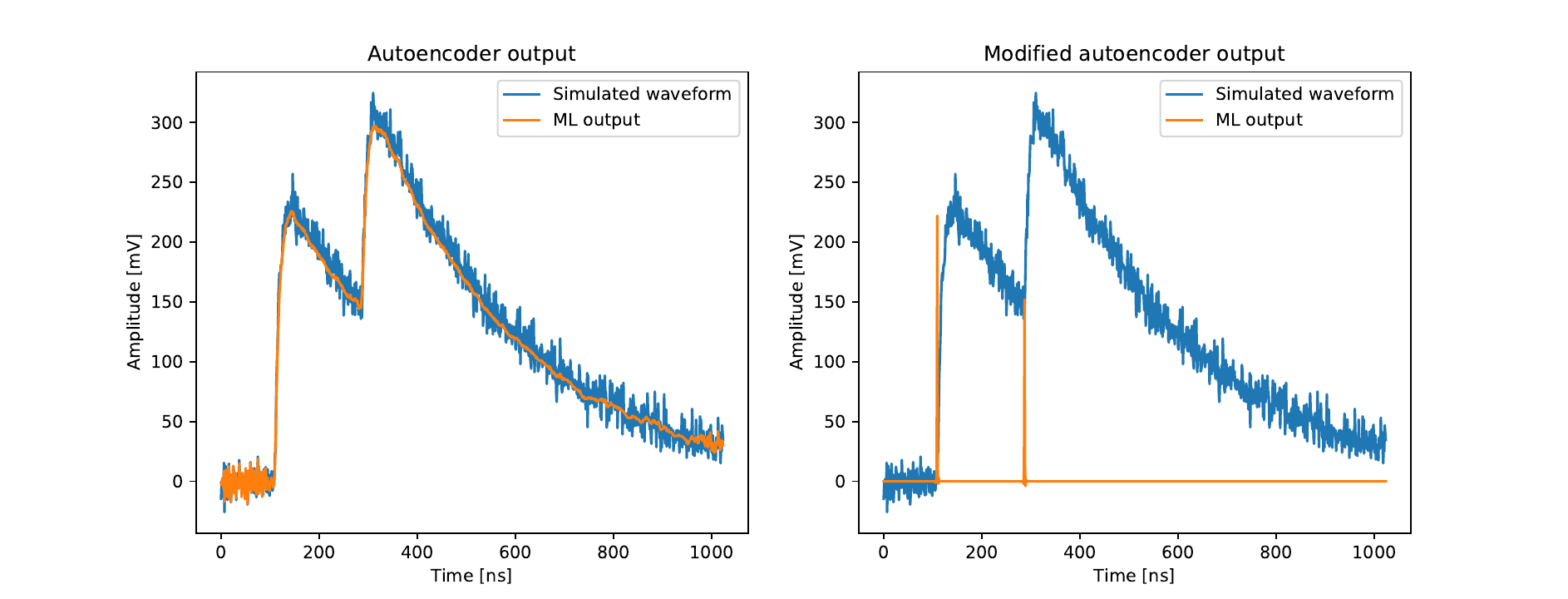}
\caption{(\textbf{Up, left:}) A simulated event with no pulses. The waveform contains only the gaussian noise. (\textbf{Up, right:}) A simulated event with two pulses, corresponding to two particles entering a crystal. (\textbf{Down, left:}) The output of applying the autoencoder network to the event from the upper right panel. The result follows the shape of the input data waveform while denoising it in the signal regions. (\textbf{Down, right:}) The result of applying the modified autoencoder to the event from the upper right panel. The result is predictions for the signal amplitude on the signal arrival positions~\cite{bib:nafski22}.
\label{fig:autoencoders}}
\end{figure}

Convolutional autoencoder neural networks~\cite{bib:CNN},\cite{bib:autoencoders} can be applied for timeseries reconstruction tasks. Such networks were applied to simulated data for the signals from scintillating BGO crystals~\cite{bib:nafski22}. The event waveforms all have a 1024~ns length and contain up to 4 individual pulses with 10~ns rise time and 300~ns fall time. The pulse arrival time is randomly assigned, following a uniform distribution and the pulse amplitude follows a gaussian distribution with 200~mV mean value and a 200~mV standard deviation. All waveforms contain gaussian noise with a 10~mV mean value as can be seen on the event with no pulses shown on the upper left panel of Figure~\ref{fig:autoencoders}. An event with two pulses from two particles entering a crystal is shown on the upper right panel of Figure~\ref{fig:autoencoders}.

The network architecture consists of an encoder composed of three 1D convolution layers followed by a decoder of three 1D transpose convolution layers. Both the encoder and decoder have one dropout layer in them. The final layer is a transpose convolution layer with a single filter, providing output that matches the input shape. Training the model on a set of 100 000 events in an unsupervised manner results in the output repeating the signal shape and denoising the data in the signal regions as shown on the lower left panel of Figure~\ref{fig:autoencoders}.

The modification to the autoencoder algorithm introduces labels for each event and uses the same architecture. The label for each event has the same length as the data with the values on all positions set to 0, except on the signal arrival positions where the value corresponds to the signal amplitude. The result of applying the modified autoencoders is shown on the lower right panel of Figure~\ref{fig:autoencoders}.

The network performance was evaluated by applying the model on an independent dataset~\cite{bib:CALOR}. The results show that individual pulses have to be more than 10~ns apart in order to be separated. The energy reconstruction shows a good correlation between the real and predicted values, however further calibration is needed. Such networks were introduced to the reconstruction software of fixed target experiments and tested on $e^{+}e^{-}\rightarrow \gamma \gamma$ annihilation events~\cite{bib:nafski23}. 

\section{Investigation of xAI techniques}

Several xAI techiques were applied to the predictions made by the modified autoencoder models: 
Integrated Gradients~\cite{bib:intgrad}, 
Vanilla Saliency~\cite{bib:vanilla}, SmoothGrad and Occlusion Sensitivity. Additionally, the output of the layers was plotted as an initial step towards the model performance visualization. All xAI techniques were tested simultaneously on three different models: M(18, lin) has a kernel size of 18 for its last filter and has a linear activation function for its output; M(14, lin) has a kernel size of 14 for its last filter and linear activation and M(18, ReLu) has a kernel size of 18 for its last filter and ReLu activation. 

\subsection{Layers output plotting}
\begin{figure}[h!]
\centering
\includegraphics[width=\columnwidth]{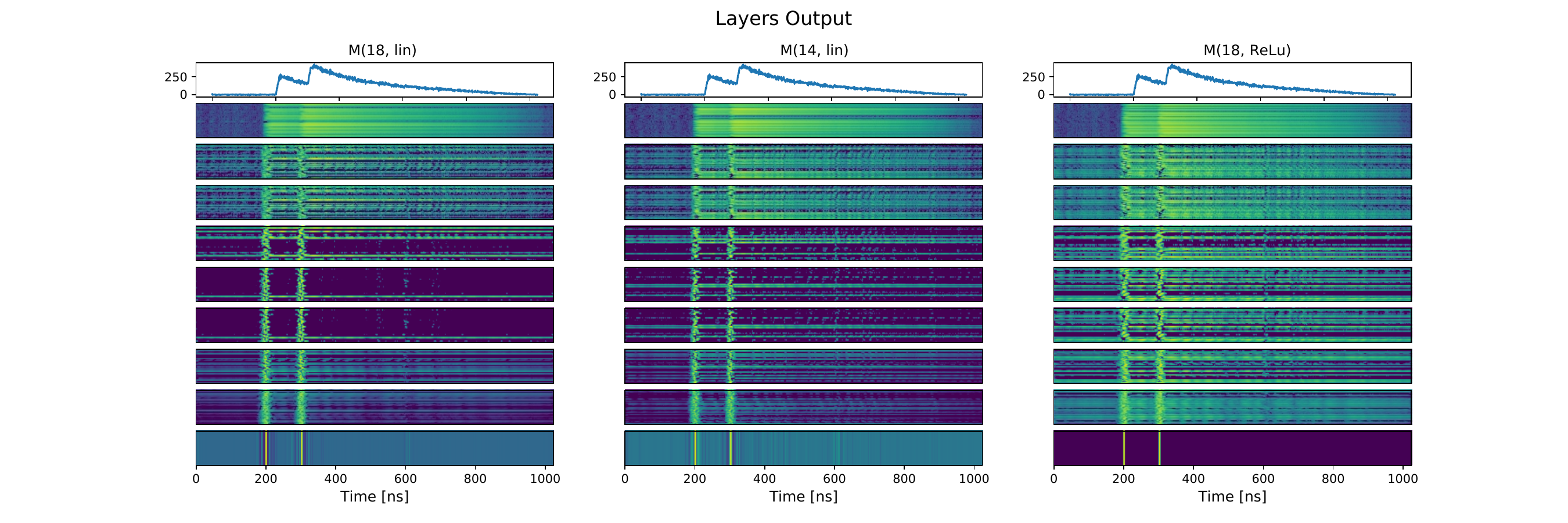}
\caption{Layers output for the three tested models for an event with two pulses created by two particles entering a crystal. The M(18, ReLu) output is clearly concentrated around the signal arrival positions while M(18, lin) and M(14, lin) have noisy output throughout the event length.
\label{fig:layers}}
\end{figure}

The first step of the explainability investigation is to plot the output of the convolution and the transpose convolution layers of the three models. The results are shown on Figure~\ref{fig:layers}. All models show similar initial filter activations. The output of the last layer for M(18, ReLu) shows positive values only around the signal arrival positions and zeroes everywhere else, while for M(18, lin) and M(14, lin), the output is noisy for the whole event length, including around the signal arrival. This result points out the significance of the last layer activation function for the correct pulse counting and arrival time determination.

\subsection{Vanilla Saliency}
\begin{figure}[h!]
\centering
\includegraphics[width=\columnwidth]{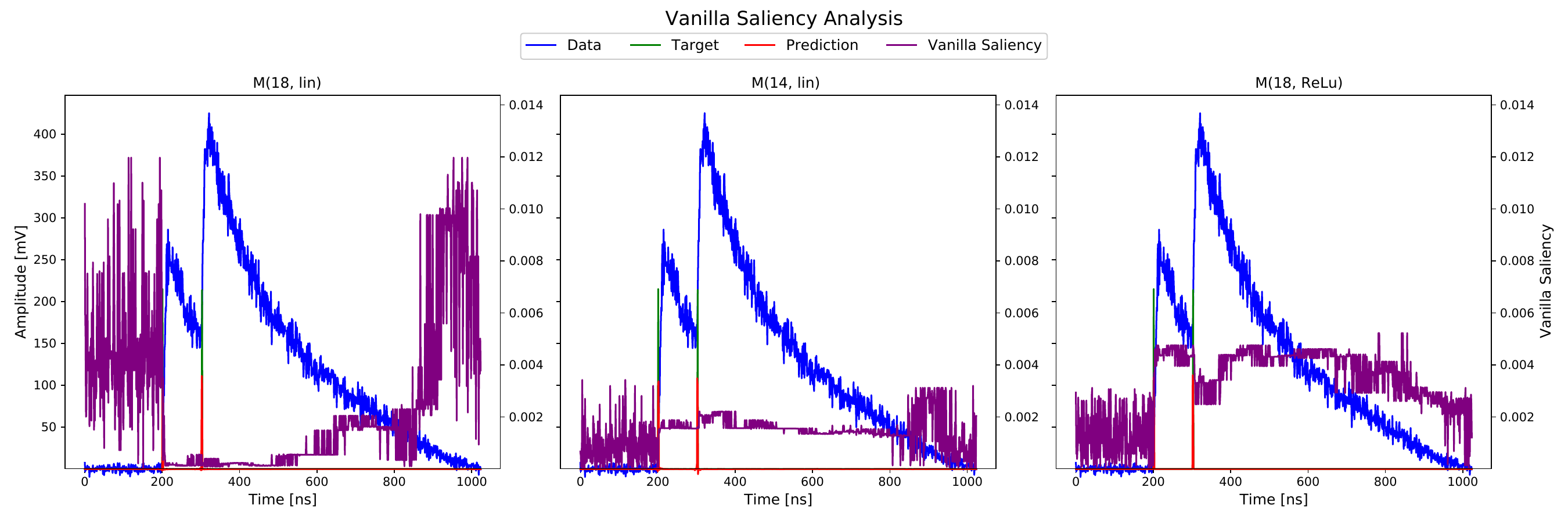}
\caption{Vanilla Saliency results for the three modified autoencoder models. All three show rapidly alternating values in the noise regions and more consistent patterns in the signal region, with gradient values lowest for M(18, lin) and highest for M(18, ReLu).
\label{fig:vanilla}}
\end{figure}

The Vanilla Saliency method calculates the gradient between the loss and the input data for each point of the data. It's a common method for explaining image classification models. The results from applying it to the modified autoencoders is shown on Figure~\ref{fig:vanilla}. The result for M(18, lin) shows very low values of the gradients in the signal regions and rapidly alternating ones in the region dominated by noise. The rapid alternation in the regions of the waveform, occupied by noise only, are also present for M(14, lin) and M(18, ReLu), yet the gradient values are higher in the signal region for M(14, lin) and even higher for M(18, ReLu).

\subsection{SmoothGrad}

The SmoothGrad method was also initially developed for explaining networks with classification purposes. 
It was suggested as an alternative to the Integrated Gradients and Vanilla Saliency methods, with the main purpose to provide less noisy output for the saliency maps. 
The goal of the method is smoothing the gradients by adding gaussian noise to the data and averaging the results in a neighborhood around a certain value as shown on Eq.\ref{eq:smoothgrad}~\cite{Smilkov2017}.
\begin{equation}
    \hat{M}_c(x) = \frac{1}{n} \sum_{1}^{n} M_c\left(x + \mathcal{N}(0,\sigma^2)\right)
    \label{eq:smoothgrad}
\end{equation}
$\mathcal{N}(0,\sigma^2)$ represents the Gaussian noise with a standard deviation $\sigma$ that is added to the data, $n$ is the size of the window in which the resulting gradients are averaged.

\begin{figure}[h!]
\centering
\includegraphics[width=\columnwidth]{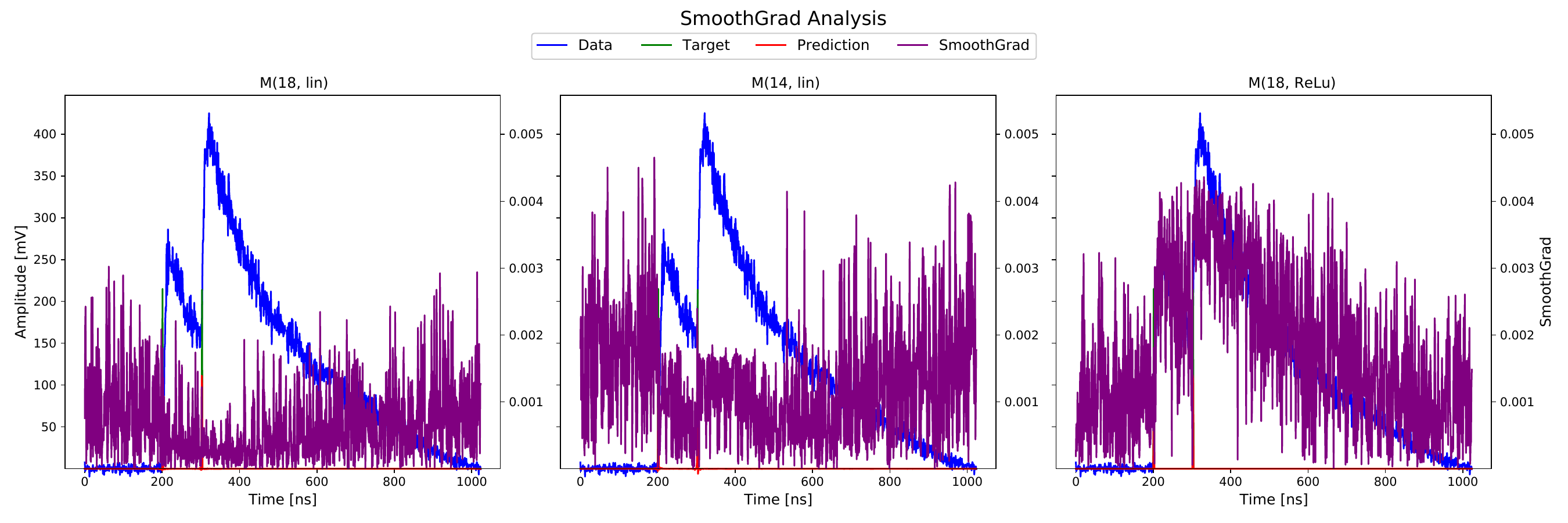}
\caption{Results from applying the SmoothGrad method to the output of the modified autoencoder models. The standard deviation $\sigma$ of the additional noise is set as half of the maximum signal value of each event. The size of the window in which the values are averaged is $n=10$. For M(18, lin) and M(14, lin), the result values are lower in the signal region, while for M(18, ReLu) they are higher.
\label{fig:smoothgrad}}
\end{figure}

The results from applying the SmoothGrad method to the modified autoencoder output are shown on Figure~\ref{fig:smoothgrad}. They show lower values in the signal regions for M(18, lin) and M(14, lin) and higher values in the signal regions for M(18, ReLu).

\subsection{Occlusion Sensitivity}

The Occlusion Sensitivity method replaces regions from the input data with a baseline, in this case set to 0, and makes predictions on the masked data. The resulting saliency maps show the values of the total loss depending on the mask position. 

\begin{figure}[h!]
\centering
\includegraphics[width=\columnwidth]{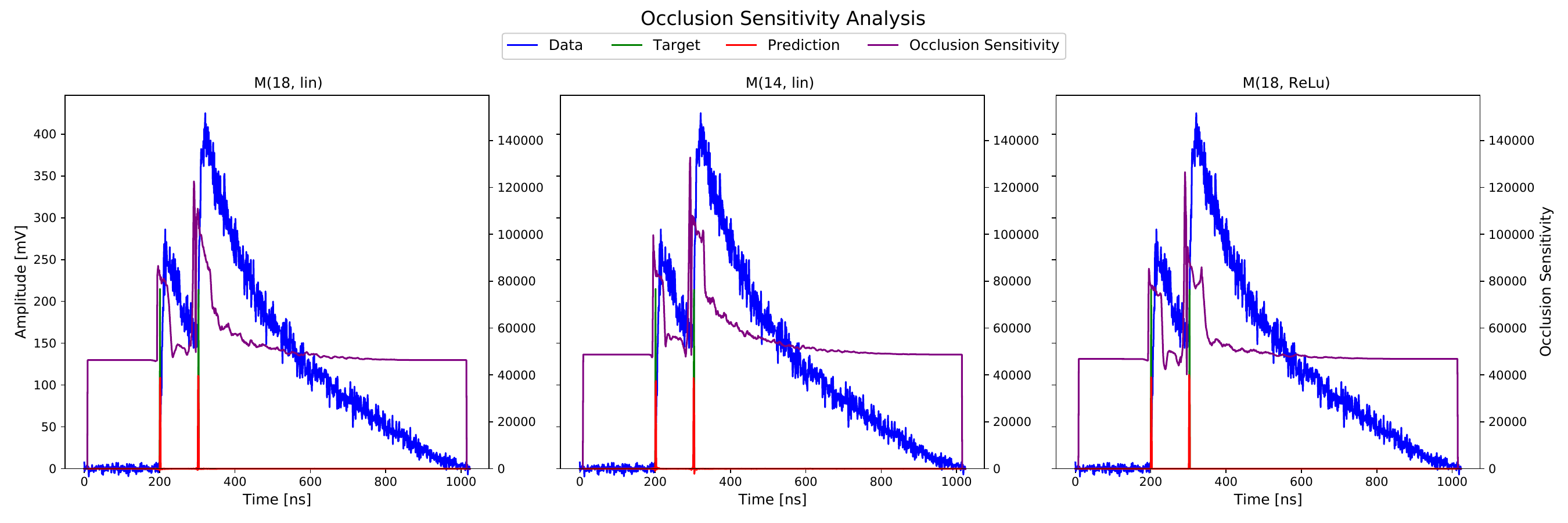}
\caption{Results from applying the Occlusion Sensitivity method to the three modified autoencoder models. The mask size that is used sets to zero 18 consecutive values. The plots show the total loss as a fuction of the mask center position. For all three models there are distinct patterns that point to the signal rise and peak regions being the most important for the signal recognition.
\label{fig:Occlusion}}
\end{figure}

The results from applying the Occlusion Sensitivity method are shown on Figure~\ref{fig:Occlusion}. For all three models, a distinct pattern of the total loss based on the mask position emerges. It shows a rapid increase in the total loss around the signal arrival, followed by a sharp decrease. For M(14, lin) and M(18, ReLu), there's a visible second peak around the signal maximum. For all three models, the total loss declines to back to the base level shortly after the maxima are reached. This points to the idea that the most influential regions from the waveforms are the signal rise and the values around its maximum. The signal decay has little influence on the overall recognition by the network.

\begin{figure}[h!]
\centering
\includegraphics[width=0.8\columnwidth]{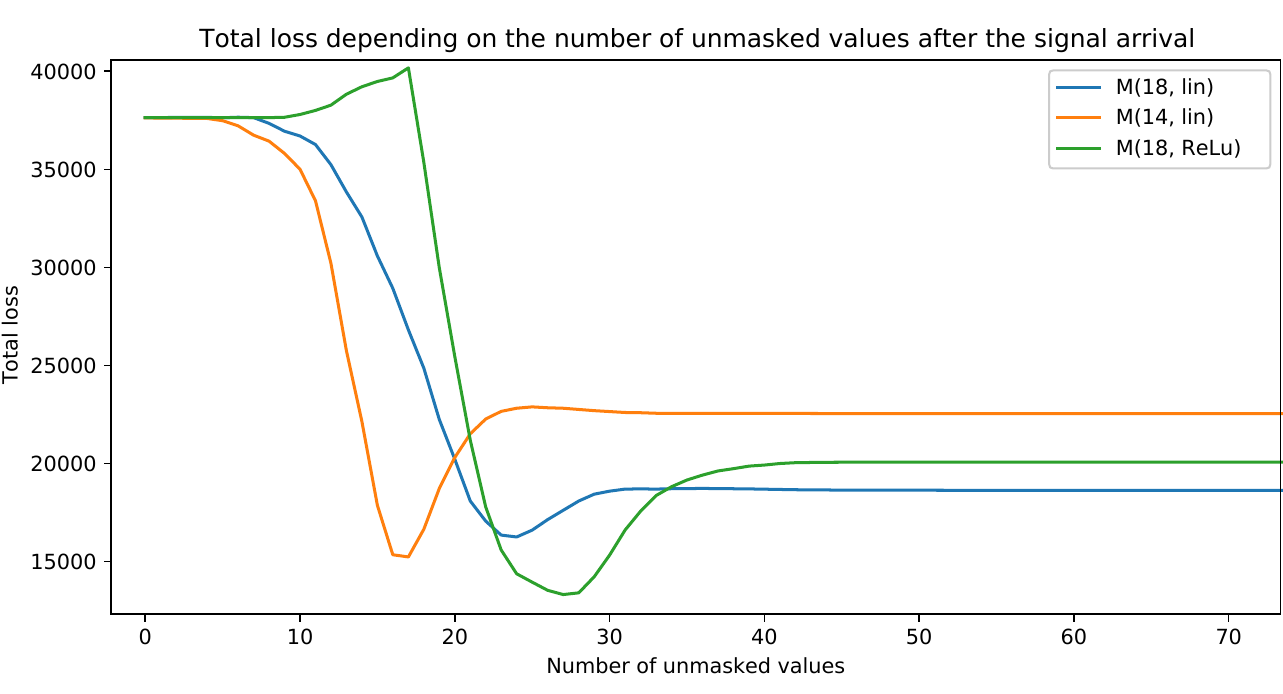}
\caption{Results from the modified occlusion sensitivity application to an event with one pulse. For all three models, a minimum in the loss is reached after unmasking less than 30 values from the beginning of the event.
\label{fig:ModOcc}}
\end{figure}

These results prompted further investigation of the method output for events where only one pulse from a single particle is present. The method was modified so that initially the whole pulse is completely masked; on each step one value after the signal arrival is unmasked and the total loss is calculated. The results are shown on Figure~\ref{fig:ModOcc}. The total loss for all three models is initially high, reaches a minimum and then returns to a constant value no matter how many more data points are unmasked. For M(14, lin), which has a smaller last layer filter kernel size, the minimum is reached earlier. M(18, ReLu) is the last one to reach a minimum in the loss, however it is the one that achieves the lowest value.

\section{Discussion and conclusions}

Modified autoencoder neural networks can be used for signal parameters extraction and data reconstruction in high energy physics detectors. A simulated dataset, matching the common signal shape in electromagnetic calorimeters with scintillating crystals was used. 
Different xAI methods were applied to the output of three models on the data. 
Plotting the layers shows the output for all filters after each layer and provides insight into how the activation function of the last layer influences the final result. 
The 
Integrated Gradients, 
Vanilla Saliency and SmoothGrad exhibit different behavior in the signal and noise regions, 
however their output is subject to further investigation. 
The Occlusion sensitivity arises as the
method providing the most understandable and further applicable information at present
stage of the presented analysis.

The results describe an analysis where the gradual unmasking of a signal allows for evaluating how effectively different models detect and separate signals. 
Initially, the total loss is very high because the signal is largely obscured and undetectable. As the signal becomes progressively unmasked,
the loss decreases, reaches a minimum, and eventually stabilizes at a constant value.

Among the models tested, M(18, ReLu) demonstrates the deepest minimum in total loss, suggesting it performs the best at detecting the signal. 
This minimum occurs when 18 values after the signal rise have been unmasked, which coincides with the kernel size of the model's last filter. 
This alignment indicates a potential relationship between the filter size and the model’s ability to accurately detect the signal. 
The combination of the signal rise time and the minimum value marks the range necessary for reliable signal detection. 
This range also provides insight into the minimum temporal separation required between two signals to distinguish them, 
which the analysis estimates to be 30 nanoseconds.

These findings suggest that M(18, ReLu) is particularly effective for tasks involving single-signal detection due to its sensitivity and precision, 
reflected in the deep minimum. The correlation between the minimum point and the filter size highlights the importance of architectural design in determining the model's receptive field. 
Interestingly, models that perform worse in single-signal detection, as indicated by shallower minima in the loss function, 
reach their minima with fewer unmasked values. 
This characteristic might make these models better suited for separating closely spaced signals, 
or double pulses, as they require less data to 
start identifying features.


Overall, this analysis provides valuable insights into how model design and loss behavior impact performance in different signal detection scenarios.

\section*{Acknowledgements} 
This work was partially supported by BNSF: KP-06-D002$\_$4/15.12.2020 within MUCCA,
CHIST-ERA-19-XAI-009.
In addition, P. Petkov and V. Kozhuharov acknowledge that partially this study is financed by the European Union-NextGenerationEU through the National Recovery and Resilience Plan of the Republic of Bulgaria, project SUMMIT BG-RRP-2.004-0008-C01. 


\end{document}